\documentclass[aps,prb,reprint,groupedaddress]{revtex4-1}
\usepackage {graphicx}  % needed for figures (not in this templat)

\usepackage{epstopdf}
\usepackage{bm}        % for math (not in this templat)
\usepackage{amssymb}   % for math (not in this templat)

\makeatletter

\newcommand{\Rmnum}[1]{\expandafter\@slowromancap\romannumeral #1@}
\makeatother

\begin{document}

\title{Electron Bubbles and the Structure of the Orbital Wavefunction}

\renewcommand{\thefootnote}{\fnsymbol{footnote}}

\author{Dohyung Ro $^{1,5}$, N. Deng$^{1,5}$,
J.D. Watson$^{1}$, M.J. Manfra$^{1,2,3}$, L.N Pfeiffer$^4$, K.W. West$^4$, and 
G.A. Cs\'{a}thy$^{1,3}$}

\affiliation{
${}^1$Department of Physics and Astronomy, Purdue University, West Lafayette, IN 47907, USA \\
${}^2$School of Materials Engineering and School of Electrical and Computer Engineering, 
          Purdue University, West Lafayette, IN 47907, USA \\
${}^3$Birck Nanotechnology Center Purdue University, West Lafayette, IN 47907, USA \\
${}^4$Department of Electrical Engineering, Princeton University, Princeton, NJ 08544, USA 
${}^5$These authors contributed equally.
}

\date{\today}

\begin{abstract}
Stripe-like and bubble-like patterns spontaneously form
in numerous physical, chemical, and biological systems
when competing long-range and short-range interactions banish uniformity.
Stripe-like and the related nematic morphology
are also under intense scrutiny in various strongly correlated electron systems.
In contrast, the electronic bubble morphology is rare. 
Some of the most intriguing electron bubbles develop in 
the two-dimensional electron gas subjected to a perpendicular magnetic field.
However, in contrast to bubbles forming in classical systems such as the 
Turing activator-inhibitor reaction or Langmuir films,
bubbles in electron gases owe their existence to elementary quantum mechanics: they are stabilized
as wavefunctions of individual electrons overlap.
Here we report a rich pattern of multi-electron bubble phases in a high Landau level and we conclude
that this richness is due to the nodal structure of the orbital component of the electronic wavefunction.
\end{abstract}

\maketitle

The pioneering Hartree-Fock theory predicted complex charge order in non-relativistic electrons residing
in the topmost Landau level of a two-dimensional electron gas (2DEG) \cite{fogler,fogler96,moessner}.
Complex charge order in 2DEGs is
manifest in the formation of stripe and bubble phases at Landau level filling factors $\nu$
at and away from half integer values, respectively.
Bubble phases are intricate electron solids:
several electrons cluster into a so called electron bubble and, in the limit of no disorder,
bubbles order on a triangular lattice to form a bubble crystal,
with a lattice constant of about three cyclotron radii \cite{fogler,fogler96}.
The Hartree-Fock theory is known to be exact for large values of the orbital Landau level index
$N \gg 1$ \cite{moessner} and it is thought to hold for $N \geq 2$ \cite{fogler,fogler96,moessner}.  
Since fluctuations are not included in the  mean-field level Hartree-Fock approach, 
the formation of bubble phases is expected to be disrupted in the $N=1$ Landau level
\cite{fogler,fogler96,moessner}.
Exact diagonalization \cite{haldane} and density matrix renormalization group (DMRG) 
studies \cite{shibata01} provided additional theoretical support for bubble formation.

Soon after the formulation of the Hartree-Fock theory, evidence for complex charge order was found in
the $N=2$ and higher Landau levels of in 2DEGs hosted in GaAs/AlGaAs \cite{lilly,du,cooper,jim}. 
Reentrant integer quantum Hall states (RIQHSs) were associated with bubble phases, while
anisotropic phases observed at half filling, termed the quantum Hall nematic,
were associated with stripe phases \cite{lilly,du,cooper,jim}.
Non-linear transport \cite{cooper,nl1}, pinning resonance \cite{lewis,lewis3}, 
onset temperature \cite{deng2}, and surface acoustic wave measurements \cite{msall,smet}
in RIQHSs also supported the bubble interpretation. 
   
 \begin{figure*}[t]
 \includegraphics[width=2\columnwidth]{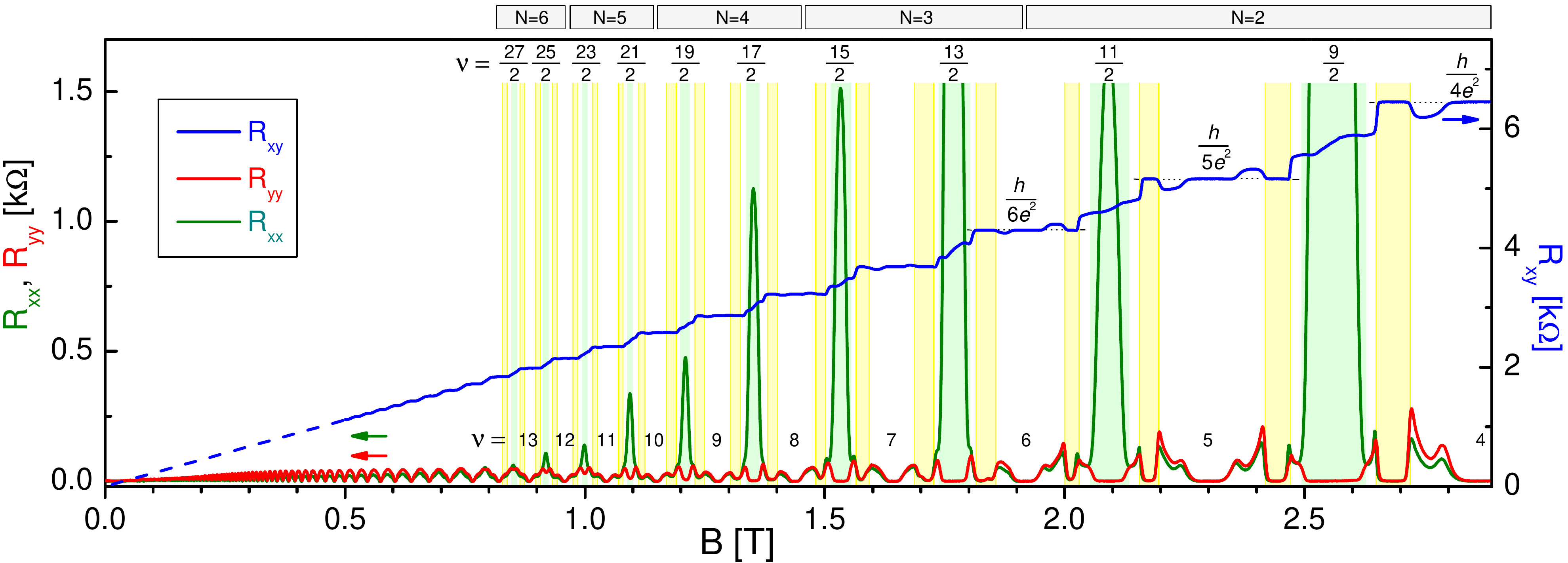}
 \caption{Dependence of magnetoresistance $R_{xx}, R_{yy}$ measured along two mutually perpendicular directions  
 and of Hall resistance $R_{xy}$  as function of the magnetic field $B$ 
 in the $N=2$ ($4<\nu<6$), $N=3$ ($6<\nu<8$),  $N=4$ ($8<\nu<10$) 
 and higher Landau levels. Here $N$ is the orbital index and $\nu$ the Landau level filling factor.
 Data were collected at $T=77$~mK.  Numbers mark the locations of integer and half-integer filling factors.
 The RIQHSs associated with bubble phases are shaded in yellow and the nematic phases in green. 
  \label{Fig1}}
 \end{figure*} 
 
Even though observations of the RIQHSs are consistent with the bubble interpretation,
they are by no means conclusive. This is because on one hand the 2DEG is deep under the surface of the GaAs crystal, 
therefore it is inaccessible to scanning probes. On the other hand,
the proliferation of RIQHSs in high Landau levels, a hallmark property of the bubble theory, 
has not been observed. Indeed, at $N \geq 2$ the Hartree-Fock
theory finds an increasing number of bubble phases as $N$ increases
\cite{fogler97,foglerBook,cote03,goerbig04,dorsey06,shibata01,yoshi02}.
A similar conclusion was reached in DMRG calculations comparing $N=2$ and $N=3$  \cite{shibata01,yoshi02}.
However, the same number of RIQHSs was seen 
both in the $N=2$ and $N=3$ Landau levels in several experiments:
in transport \cite{lilly,du,smet,n2,n3,zudov}, in microwave pinning resonance \cite{lewis,lewis3},
and in surface acoustic wave attenuation measurements \cite{smet,n3}. 
While it is possible that a single RIQHS is associated with multiple bubble phases, in lack of
direct probes of the morphology such an assumption could not be tested.
A lack of observation of the proliferation of bubble phases at large Landau indices                       
is particularly unsettling since the Hartree-Fock theory is 
expected to provide an increasingly better description of the bubbles 
as $N$ increases \cite{moessner}.

In this Rapid Communication  we study magnetotransport of RIQHSs in high Landau levels of an exceptional quality 
2DEG confined to GaAs/AlGaAs. 
Consistent with earlier results, in the $N=2$ Landau level we observe four RIQHSs
which, based on symmetry considerations, were associated with a single type of bubble phase.
In contrast, we find that in the $N=3$ Landau level the family of RIQHSs is richer than previously reported.
Here we find evidence for eight distinct RIQHSs. Our results indicate that,
when disorder is sufficiently low, RIQHSs proliferate in a high Landau level.
Such a proliferation of the RIQHSs so far was a missing element of the bubble interpretation
of the RIQHSs; it therefore                 significantly strengthens 
the bubble interpretation and
allows us to identify the types of bubbles forming in the $N=2$ and $3$ Landau levels. 
Furthermore, our results highlight a fundamental difference between classical and quantum bubbles.
We found evidence that the structure of the electronic wavefunction determines the richness of multi-electron bubbles:
when the wavefunction has multiple nodal lines, multiple bubble phases form, each with a different 
number of electrons per bubble.

In Fig.\ref{Fig1} we plot representative magnetotransport traces in the $N=2$ and higher Landau levels.
$R_{xx}$ and $R_{yy}$ are the longitudinal magnetoresistances,
as measured along two mutually perpendicular crystal axes of GaAs, and $R_{xy}$ is the Hall resistance.
Our sample has a density $n=2.8 \times 10^{11}$cm$^{-2}$ and mobility
$15 \times 10^6$cm$^2$/Vs. Further details on this sample and our measurement
setup can be found in Ref.\cite{deng2}.
Vanishing magnetoresistances $R_{xx} = R_{yy}=0$ and Hall resistance quantized to $R_{xy}=h/ie^2$
at integer Landau filling factors $\nu=i$, with $i=4,5,6,...$, signal integer quantum Hall states (IQHS) \cite{klitzing}.
The striking strong resistance anisotropies near half-filling, at $\nu=9/2, 11/2, ..., 27/2$
indicate the quantum Hall nematic \cite{lilly,du}, phases
related to the stripes of the Hartree-Fock theory \cite{fogler,fogler96,moessner,frad0,frad1,radzi}. 
Resistance anisotropy in Landau levels as high as $N=6$ is a fingerprint of a remarkably high quality sample. 

Based on prior work, transport behavior near $\nu=4$ may be understood as follows. 
At $\nu=4$, which occurs at $B=2.90$~T in our sample,
the $N=0$ and $N=1$ Landau levels are full and the $N=2$ Landau level is empty. By lowering 
$B$ we populate the $N=2$ Landau level and therefore
increase of $n^*$, the areal density of electron quasiparticles in the topmost Landau level.
At low $n^*$ disorder localizes the electron quasiparticles, hence an
integer plateau $R_{xx}=0$ and $R_{xy}=h/4e^2$ develops in Fig.\ref{Fig1} between
$B=2.8$~T and $B=2.90$~T. 
As $n^*$ increases, the Coulomb energy of the electrons overcomes
disorder effects, leading to a crossover to a Wigner solid \cite{ws1}.
The Wigner solid cannot be distinguished by dc transport, but its signatures were seen in microwave
resonance \cite{ws1,ws11}, compressibility \cite{ws2}, resistively detected NMR \cite{ws3,ws33}, and
tunneling measurements \cite{ws4}.
A further increase in $n^*$ leads to the observation of reentrance:
first there is a deviation from integer quantization, then there is a conspicuous return
to quantization \cite{lilly,du,cooper}. 
Such a behavior is seen in the regions near $B=2.44$ and $2.68~T$ and other regions shaded in yellow in Fig.\ref{Fig1}.
Since an Anderson insulator is not expected to be reentrant, reentrance signals a collective insulator
such as the electronic bubble phase \cite{du,cooper}.

Consistent with earlier results, in the $N=2$ Landau level we observe four RIQHSs
centered at $\nu=4+0.29$, $5-0.29$, $5+0.29$, and $6-0.29$ \cite{du,cooper,deng2}. 
These are seen in Fig.\ref{Fig1}.
The four filling factors are related by particle-hole symmetry, hence these four RIQHSs 
are associated with a single type of bubble phase.
Calculations at these filling factors found a bubble phase with $M=2$ electrons per bubble
\cite{fogler,fogler96,haldane,shibata01,cote03,goerbig04,yoshi02,dorsey06}. 

\begin{figure}[t]
 \includegraphics[width=.9\columnwidth]{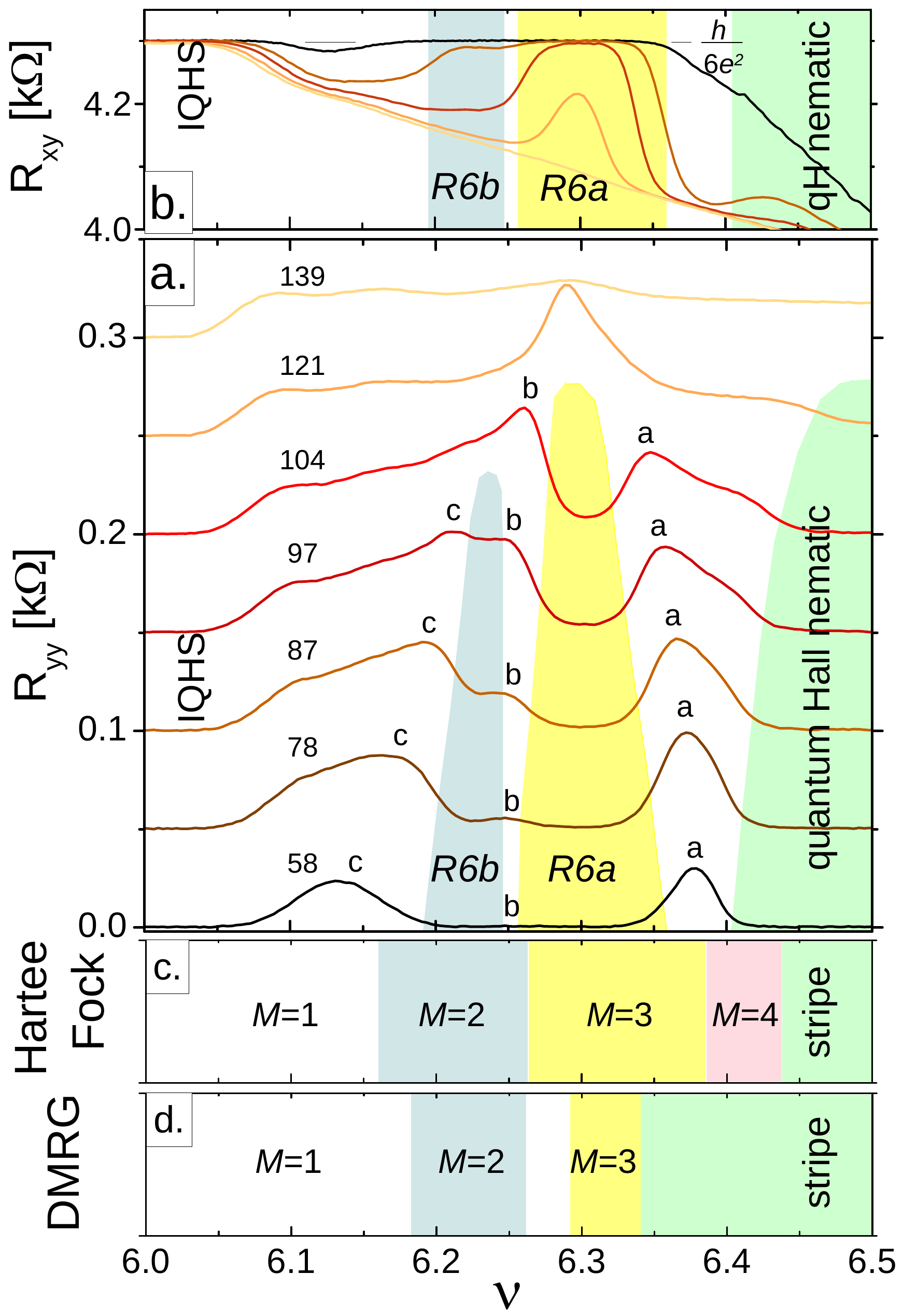}
 \caption{Panel $a$: Waterfall plot of the temperature evolution of the 
 magnetoresistane in the $N=3$ LL, for $6 < \nu < 6.5$. 
 Traces are shifted by $50$~$\Omega$ one relative to the other.
 Local maximum marked $b$ delimits the two distinct RIQHSs  
 labeled $R6a$ and $R6b$ and which are shaded in yellow and blue, respectively.
 Panel $b$: Hall resistance in the region of the reentrant phases. 
 Traces at $T=78$ and $97$~mK were omitted to reduce clutter.  Panels $c$ and $d$
 show the phase diagrams of competing bubble phases in the $N=3$ Landau level, as obtained
 using Hartree-Fock \cite{yoshi02,cote03,goerbig04}  
 and DMRG calculations \cite{yoshi02}, respectively. Bubble phases with two, three, and four 
 electrons per bubble are shaded in blue, yellow, and pink, respectively.
 \label{Fig2}}
 \end{figure}

To our surprise, the magnetoresistance in the region of reentrance
near $B=1.83$~T or near $\nu=6.3$ in Fig.\ref{Fig1} appears wider than expected and it reveals an unfamiliar structure
in the $N=3$ Landau level. We proceed to examine details of this structure more closely. 
The $T=104$~mK traces in Fig.\ref{Fig2} exhibit a vanishing magnetoresistance $R_{yy}$ and a quantized
Hall resistance $R_{xy}=h/6e^2$ at $\nu=6.30$, 
signaling therefore the development of a RIQHS labeled $R6a$. The local minimum of the magnetoresistance
is found between two local maxima labeled $a$ and $b$, in the region shaded in yellow in Fig.\ref{Fig2}a. 
However, in addition to the local minimum in $R_{yy}$ at $\nu=6.30$, a second local minimum appears
at $\nu=6.23$ as the temperature is lowered to $T=97$~mK; this second minimum is
located between the local maxima labeled $b$ and $c$ in Fig.\ref{Fig2}a, in a region shaded in blue.
As the temperature is further lowered to $T=58$~mK, $R_{yy}$ at this second local minimum vanishes,
while the Hall resistance is quantized to $R_{xy}=h/6e^2$.
This data unveils a new RIQHS at $\nu=6.23$, labeled $R6b$ in Fig.\ref{Fig2}a.
Because of the resistive feature $b$, this new RIQHS is distinct from $R6a$.
We thus observed a doubling of reentrance of the $\nu=6$ IQHS           
in the $6 < \nu < 6.5$ range of filling factors of the $N=3$ Landau level:
instead of a single RIQHS, in this region we encounter two distinct RIQHSs.
Therefore the number of RIQHSs in the $N=3$ Landau level
exceeds that in the $N=2$ Landau level, a hallmark prediction of theory of the bubbles.
Such a doubling of the RIQHSs can be readily interpreted as 
being due to two distinct multi-electron bubble phases.
Our data provides a first indication of the proliferation of multi-electron bubble phases in a high Landau level
and offers a strong, albeit indirect evidence for the bubble interpretation of the RIQHSs.

 \begin{figure}[t]
 \includegraphics[width=0.95\columnwidth]{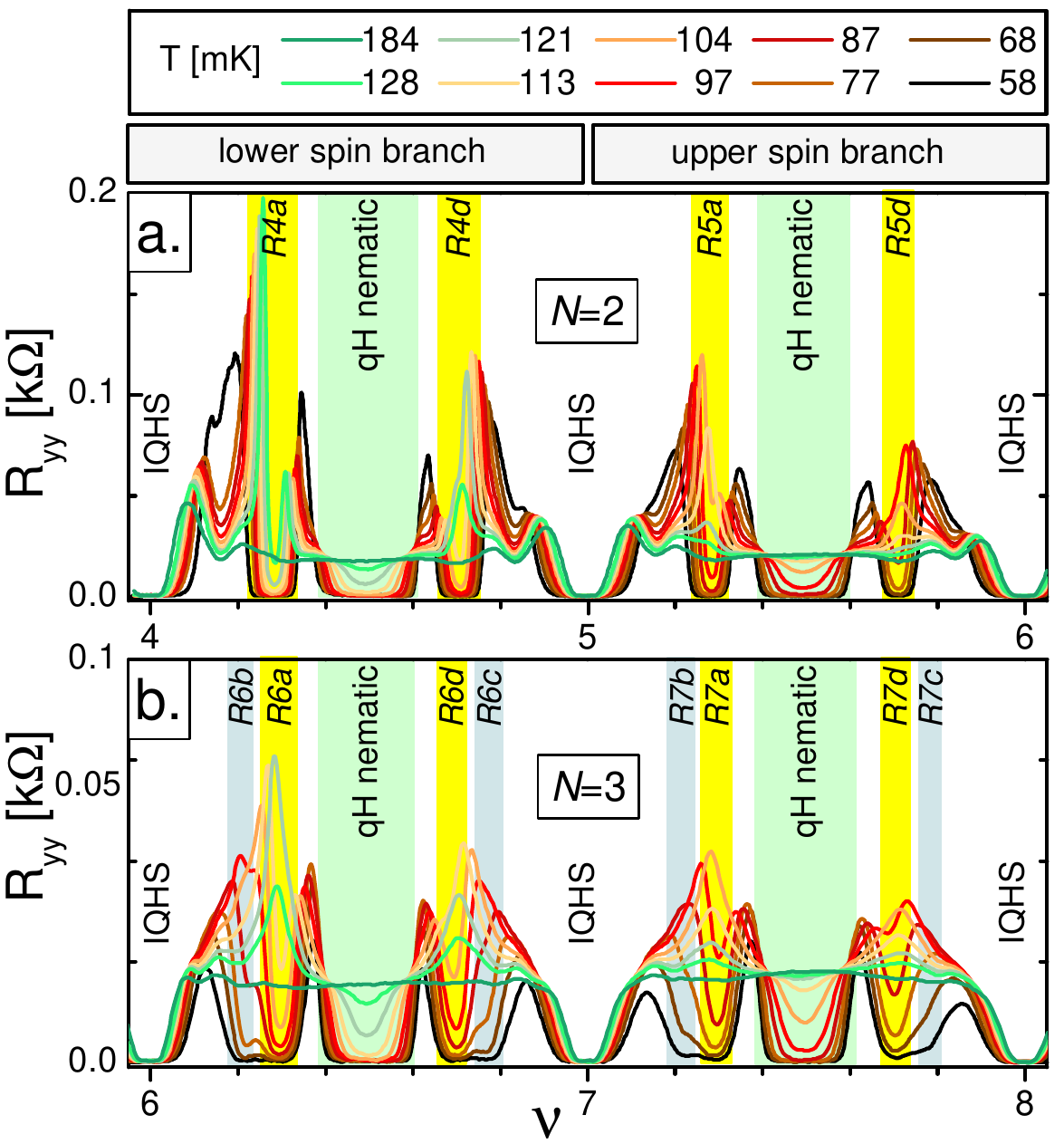}
 \caption{Evolution with temperature of magnetoresistance in the $N=2$ (panel a.) and $N=3$ 
 (panel b.) Landau levels. 
 Note the different vertical scales. Numbers in the legend shown on top are temperatures in units of mK.
 The two spin branches are also marked.
 RIQHSs are shaded in yellow and blue, stripe phases are shaded in green, while the IQHSs are unshaded.
 In the $N=3$ Landau level the resistive feature observed at the lowest temperatures delimits
 the distinct RIQHSs.
  \label{Fig3}}
 \end{figure}
 
Our data contain additional details that further strengthen the bubble interpretation of RIQHSs.
First, a peculiar feature of the Hartree-Fock \cite{fogler97,foglerBook,cote03,goerbig04,dorsey06,yoshi02} 
and of DMRG calculations \cite{shibata01,yoshi02}
is that the multiple bubble phases involved must be close in energy. Our data in Fig.\ref{Fig2} shows this is
the case for the RIQHSs $R6a$ and $R6b$.
We estimate that the two onset temperatures are within 15\% of each other.
Second, in accordance with the Hartree-Fock theory,
the doubling of reentrance is an orbitally driven effect. Indeed, data in Fig.\ref{Fig3}b
show that both spin branches of the $N=3$ Landau level exhibit doubling of the reentrant behavior.
This means that the physics is independent of the spin quantum number and therefore
the RIQHSs we observe, including the newly seen $R6b$, $R6c$, $R7b$, and $R7c$,
precipitate because of orbital effects. These groups of RIQHSs form at filling factors related by
particle-hole symmetry. 
 
Using the results of the Hartree-Fock and DMRG calculations, we can assign bubble phases to the RIQHSs.
Fig.\ref{Fig2}c and Fig.\ref{Fig2}d displays the predictions for ground states in the $N=3$ Landau level
of the Hartree-Fock \cite{yoshi02,cote03,goerbig04} and DMRG \cite{yoshi02} calculations, respectively. 
Our RIQHSs are in excellent alignment with the bubble phases with $M=2$ and $M=3$ electrons per bubble.
Furthermore, the range of stability for the $M=1$ bubbles or the Wigner solid
overlaps with, but it is considerably wider than the measured $\nu=6$ integer quantum Hall plateau.
We conclude that the RIQHS labeled $R6a$ centered around $\nu=6.30$ is a bubble phase with $M=3$ electrons
per bubble, whereas the slightly weaker $R6b$ centered around $\nu=6.23$ is a $M=2$ bubble phase.
We see no evidence for the bubbles with $M=4$ electrons of the Hartree-Fock theory \cite{yoshi02,cote03,goerbig04};
at the expected filling factors we instead observe the quantum Hall nematic phase. DMRG calculations
capture the transition from the $R6a$ bubble to the nematic more accurately.
By the same token, the RIQHSs in the $N=2$ Landau level, shown in Fig.\ref{Fig3}a, 
are associated with $M=2$ bubble phases.

Data from Fig.\ref{Fig2}a show that the onset temperature of $R6a$, the RIQHS closest to half filling,
exceeds that of $R6b$. These energy scales should compare favorably to cohesive energies 
of the $M=3$ and $M=2$ bubble phases calculated within the Hartree-Fock approach. 
Our data qualitatively agree with results from Refs.\cite{cote03,dorsey06}, reporting
the most stable $M=3$ bubbles forming closest to half filling. However,
our data compare unfavorably to results from Refs.\cite{fogler97,goerbig04}.
The contradictory results of cohesive energy calculations cannot be accounted
by the different dielectric functions used: Refs.\cite{fogler97,dorsey06} 
include screening through a wavenumber-dependent dielectric function, whereas
Refs.\cite{cote03,goerbig04} use a constant dielectric function.
Authors of Ref.\cite{dorsey06}, however, find that the finite thickness of 
the electron layer has a strong influence on cohesive energies. These results suggest
that finite thickness effects or other effects that modify the
short-range part of the electron interaction have a strong impact on cohesive energies.
The contradictory results on cohesive energy calculations, however, do not alter the basic 
observation of two RIQHSs in $N=3$ and the assignment of these RIQHSs to multi-electron bubble phases.

 \begin{figure}[t]
 \includegraphics[width=0.8\columnwidth]{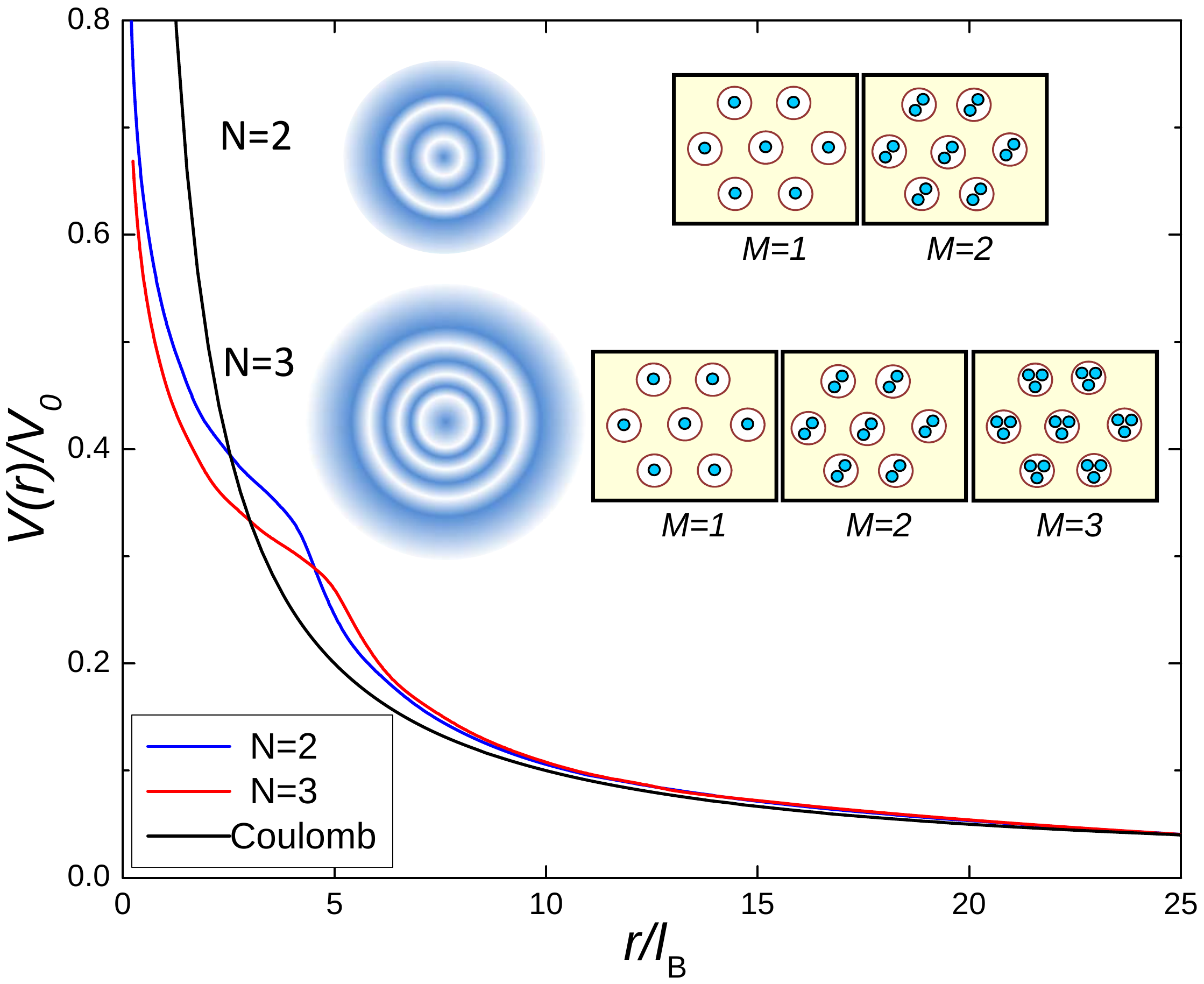}
 \caption{The interaction energy for $N=2$, $N=3$, and the Coulomb expression, 
 in units of $V_0=e^2/(4 \pi \epsilon l_B)$. When the wavefunctions overlap,
 energies deviate one from another and also from the Coulomb expression.
 The inset correlates the structure of the wavefunction $|\psi|^2$ in the symmetric gauge
 with the types of bubble phases allowed.
 At $N=2$ the wavefunction has two nodes and there are two types of bubble phases, with $M=1$ and $M=2$.
 In contrast, at $N=3$ the wavefunction has three nodes and there are three different bubble phases.
  \label{Fig4}}
 \end{figure}
 
At the origin of bubbles one finds competing short-range and long-range interactions.
For electronic bubbles the long-range interaction is due to the Coulomb repulsion, while 
the short-range interaction is due to overlapping of the quantum mechanical 
wavefunctions $\psi$ of the electrons \cite{fogler,fogler96,moessner}.
In Fig.\ref{Fig4} we show a representation of $|\psi|^2$ for $N=2$ and $N=3$
in the symmetric gauge. The overlap of these objects when their geometric centers
are close generates the bubbles. 
Because of the Landau index dependent structure of the wavefunction,
the short-range interaction also depends on the Landau index.
However, wavefunctions $\psi$ depend on the choice
of the gauge potential describing the magnetic field \cite{ezawa}. 
In contrast, observables are gauge-independent. We suggest that the gauge-independent quantity relevant
for the formation of the bubbles is the number of nodal lines of the electronic wavefunction.
It was already known that nodes in the wavefunction play a role in stabilizing the bubbles \cite{jim}.
Our discovery of double reentrance 
in the $N=3$ Landau level brought to the fore the profound effects of multiple nodal lines
in stabilizing a rich set of multi-electron bubbles. Nodal lines in Fig.\ref{Fig4} are shown as white circles.
Our results show that, when the Wigner solid is included as the $M=1$ electron bubble,
the number of bubble phases for $N=2$ and $N=3$
coincides with the number of nodal lines and with the Landau index $N$.
In contrast to bubble phases, quantum Hall
nematics in our experiment did not exhibit any obvious
dependence on the nodal structure of the wavefunction.

Our findings highlight fundamental differences between classical 
and quantum mechanical bubbles. Examples for the former are bubbles
in the Turing activator-inhibitor system or in Langmuir films \cite{ball}.
In classical systems there is only one type of bubble; changing the density of the system
often results in a change of the size of the bubbles. 
In contrast, there are different types of quantum bubbles are allowed in the 2DEG;
as the density of the quasiparticles is
changed, there will be either a phase transition or a crossover between 
different types of bubble phases. 

We used the weak but distinctive resistive feature marked $b$ in Fig.\ref{Fig2}a to
identify the RIQHSs labeled $R6a$ and $R6b$. It is interesting to note that
within the Hartree-Fock theory, a sharp phase transition between the two
bubble phases is expected \cite{fogler97,foglerBook,cote03,goerbig04,dorsey06,shibata01,yoshi02}.
A resistive feature of finite width may be the result
of rounding due to the presence of disorder in our sample, which is not included in the theory.
A resistive feature may also be caused by a backscattering channel provided by percolating 
paths between coexisting domains of the two bubble phases \cite{foglerBook}.
It is interesting to note that the uncolored region between the $M=2$ and $M=3$ bubble phases in Fig.\ref{Fig2}d
at which DMRG calculations could not identify the ground state \cite{yoshi02}
has a very good overlap with the resistive feature we observe.

Recent observations of RIQHSs in graphene, a Dirac material, offers the opportunity
to study electron bubbles in a new platform \cite{dean}. Results underscore
the host-independent aspects of the physics at play
and offer the chance to study novel effects, such as the dependence of the RIQHSs on the
valley degree of freedom.

To conclude, in the $N=3$ Landau level of a high quality 2DEG confined to GaAs/AlGaAs we observed
a double reentrance of the integer quantum Hall effect. Observations provide evidence
for the proliferation of the RIQHSs in high Landau levels and therefore lend a strong support to the
bubble interpretation of the RIQHSs. Our result highlight the role of quantum mechanics in forming the bubbles.
In particular, the richness of the bubble phases in the $N=3$ Landau level is attributed to the
presence of three nodal lines in the electronic wavefunction. 

{\it Note.} Observations from this manuscript were reported earlier \cite{kevin} and
related observations in a sample with alloy disorder were recently published by Fu et al. \cite{zud}.

Measurements at Purdue were supported in part by the NSF grant DMR 1505866. The sample growth effort of M.J.M.
was supported by the DOE BES award DE-SC0006671, while that of L.N.P. and K.W.W. by the Gordon
and Betty Moore Foundation Grant No. GBMF 4420, and the NSF MRSEC Grant No. DMR-1420541.

\end{document}